\documentclass[lettersize,journal]{IEEEtran}
\usepackage{amsmath,amsfonts,amssymb,amsthm}
\usepackage{dsfont} 
\usepackage{array}
\usepackage[caption=false,font=normalsize,labelfont=sf,textfont=sf]{subfig}
\usepackage{textcomp}
\usepackage{stfloats}
\usepackage{url}
\usepackage{verbatim}
\usepackage{graphicx}
\usepackage{float}
\def\BibTeX{{\rm B\kern-.05em{\sc i\kern-.025em b}\kern-.08em
    T\kern-.1667em\lower.7ex\hbox{E}\kern-.125emX}}
\usepackage{balance}
\usepackage{color}
\usepackage{cite}
\usepackage{algorithm}
\usepackage{algorithmicx}
\usepackage[noend]{algpseudocode}
\usepackage{booktabs}
\theoremstyle{plain}
\newtheorem{theorem}{Theorem}
\newtheorem{lemma}{Lemma}

\newtheorem{remark}{Remark}
\usepackage{orcidlink,hyperref}
\hypersetup{hidelinks,breaklinks=true}
\usepackage{epstopdf}

\begin{document}
\title{Timely and Energy-Efficient Information Delivery in Heterogeneous Correlated Random Access Networks}
\author{Anshan Yuan\hspace{0.5mm}, Xinghua Sun\hspace{0.5mm},~\IEEEmembership{Member,~IEEE},  Yayu Gao\hspace{0.5mm},~\IEEEmembership{Member,~IEEE}, Wen Zhan\hspace{0.5mm},~\IEEEmembership{Member,~IEEE},\\and Xiang Chen\hspace{0.5mm},~\IEEEmembership{Member,~IEEE}
\thanks{
Anshan Yuan, Xinghua Sun, and Wen Zhan are with the School of Electronics and Communication Engineering, Shenzhen Campus of Sun Yat-sen University, Shenzhen 518107, China (e-mail: \url{yuanansh@mail2.sysu.edu.cn}; \url{sunxinghua@mail.sysu.edu.cn}; \url{zhanw6@mail.sysu.edu.cn}).

Yayu Gao is with the School of Electronic Information and Communications, Huazhong University of Science and Technology, Wuhan 430074, China (e-mail: \url{yayugao@hust.edu.cn}).

Xiang Chen is with the School of Electronics and Information Technology, Sun Yat-sen University, Guangzhou, China (e-mail: \url{chenxiang@mail.sysu.edu.cn}).
}
}


\maketitle

\begin{abstract}

This paper characterizes and jointly optimizes Age of Information (AoI) and energy efficiency in heterogeneous correlated random access networks, where each sensor adopts a distinct transmission probability and its observations are correlated with those of other sensors. An analytical model is proposed to analyze AoI and energy efficiency for each sensor. Closed-form expressions for long-term average AoI and energy efficiency are derived, explicitly accounting for spatial correlation and state-dependent power consumption. 
By constraining sensors to adopt the same transmission probability, three unified transmission strategies are derived: the age-optimal strategy ($q_A^*$), the energy-efficiency optimal strategy ($q_E^*$), and the Pareto-optimal strategy ($q^*$), which jointly optimizes AoI and energy efficiency. A bounded exhaustive search with $\mathcal{O}(\frac{1}{nq_{\epsilon}})$ complexity guarantees efficient computation of $q^*$.
Theoretically, the correlation gain is proven to significantly enhance both metrics under spatial correlation.
To exploit sensor heterogeneity, a gradient-based iterative algorithm, Multi-Start Projected Adaptive Moment Estimation (MS-PAdam), is proposed to jointly optimize all sensors' transmission probabilities, efficiently converging to the optimal AoI–energy-efficiency tradeoff. Crucially, MS-PAdam adaptively suppresses transmissions where marginal gains are outweighed by correlated neighbors' contributions, substantially alleviating competition. Numerical results show MS-PAdam outperforms unified strategies, achieving harmonious operation that mitigates AoI/energy degradation in contention-intensive scenarios.
\end{abstract}


\section{Introduction}
The evolving wireless ecosystem and the rapid expansion of Machine-Type Devices (MTDs) have significantly increased the prevalence of machine-to-machine (M2M) communications, enabling seamless integration of the physical and information worlds. This has led to applications such as environmental monitoring, intelligent manufacturing, and real-time autonomous systems \cite{AnthonyEphremides_AoI, yu2022age, 5G_Advanced_JSAC, CaoJie_Wirel_Comm}. Within these applications, MTDs are typically battery-operated and lack recharging capability. Moreover, they must collect timely environmental status information through sensing and transmission under finite energy constraints, thereby facilitating energy-efficient real-time control and precise decision-making. 
In response, Age of Information (AoI) \cite{kaul2012AoI} and energy efficiency have emerged as critical performance metrics in the Energy-Aware Internet of Things (EA-IoT). AoI is defined as the time elapsed since the most recent packet received at the base station was generated by a sensor, whereas energy efficiency quantifies the effective utilization of a device’s energy over its lifetime. In EA-IoT, sensors operate with limited energy to update environmental status information, necessitating both favorable AoI and energy-efficiency performance. Random access, valued for its scalability, simplicity, and elegance, has become the most widely adopted multiple access scheme in EA-IoT. However, the AoI and energy efficiency of MTDs in random access can be severely degraded by frequent collisions resulting from the distributed nature of sensor access, which underscores the importance of addressing an energy-efficient access strategy associated with the growing number of distributed sensors.
\subsection{AoI Analysis in Random Access Networks}\label{subsec: AoI in RA}

Extensive research has employed AoI as the performance metric for monitoring and control systems in multiple access networks. While centralized scheduling algorithms achieve better AoI performance, they incur substantial signaling overhead due to their reliance on a central controller. In contrast, random access enables distributed channel contention, making it more scalable for EA-IoT deployments.
In the link pair interference scenario, where multiple transmitter-receiver pairs coexist and compete for the channel under a random access scheme, the studies in \cite{yang2021understanding} and \cite{yang2022Spatiotemporal} analyzed and optimized AoI. Similarly, \cite{2022fangmingIoTJ} and \cite{2022fangmingTGCN} established a fixed-point equation for the successful transmission probability of nodes and further analyzed and optimized network AoI. 
Recently, \cite{2024fangmingTWC} extended the works in \cite{2022fangmingIoTJ} and investigated AoI of the Aloha networks with an age threshold, where each transmitter accesses the channel when its AoI is larger than a specific threshold.

Unlike the link pair interference scenario, the following studies consider a scenario with multiple sensors and a single receiver in the random access network. \cite{37} investigated a stochastic random access protocol based on the Whittle Index and provided the steady-state distribution of achievable AoI. \cite{38} examined the average AoI under the Carrier Sense Multiple Access protocol and optimized AoI by tuning the backoff duration.
\cite{40} analyzed the AoI and obtained an approximately optimal transmission probability for nodes in slotted Aloha networks. 
Based on the analytical model proposed in \cite{zhanwen2018TWC}, \cite{2023deweiTNSE} derived the explicit expression of the Peak AoI and further jointly optimized the packet arrival rate and access probability in Aloha networks. The optimization results reveal that the exponential growth of the Peak AoI with the increase in the number of network nodes can be reduced to linear growth. 
\cite{41} considered slotted Aloha networks with an age threshold and provided expressions for the average AoI in the network. Based on the threshold Aloha strategy proposed in \cite{41}, \cite{42} further analyzed the relationship between the optimal average AoI and network scale. Under random sampling, \cite{43} introduced the concept of AoI gain to quantify the reduction in the receiver's instantaneous AoI following a successful packet transmission. It was further suggested that selective packet transmission based on AoI gain at the sender can effectively optimize AoI. \cite{44} presented the lower bound of AoI in framed slotted Aloha networks and studied the impacts of packet arrival and transmission probabilities.

\subsection{AoI Analysis with Correlated Status Updates}\label{subsec: AoI in correlated}

The aforementioned studies have comprehensively analyzed and optimized the AoI in random access networks, addressing both theoretical modeling and algorithm design. However, all of these works only explored the AoI in random access scenarios from a temporal perspective, without considering the spatial correlation of nodes in intelligent IoT scenarios, i.e., \textit{each sensor is independent or uncorrelated}, thus the AoI of each sensor is only influenced by the successful channel access of itself. 
As such, the aforementioned studies typically aim to optimize either the AoI of individual sensors or their weighted sum as the network AoI.

This is not entirely accurate in certain IoT applications, where many monitoring and control tasks require information from correlated sources. For instance, in a smart camera network, multiple cameras with overlapping fields of view may monitor a shared scene where a remote base station can infer the state information of several cameras using the image captured by just one camera. Similar examples include wireless sensor networks where sensors collect spatially correlated updates as they probably observe the same physical process. In these scenarios, a status update from one sensor often carries information about the current status of other sensors.
Recently, \cite{zhangheng2021TWC, hribar2019IOTJ, hribar2022IOTJ} have demonstrated that environmental status updates collected by sensors exhibit spatial correlation. This correlation is primarily reflected in the similarity of state information observed by adjacent nodes and in the ability to estimate the environment state of a continuous region using sampling data from multiple discrete nodes. Leveraging this spatial correlation allows for the adjustment of the transmission parameter while preserving perceptual accuracy.

Some studies have investigated the AoI in the correlated networks. 
\cite{Qing2019Ton} and \cite{Zhoubo2020Globecom} considered a network of cameras with overlapping fields-of-view and formulated an optimization problem based on AoI to study processing and scheduling in this correlated setting.
For monitoring systems with common observations, \cite{Anders2019WCL} proposed scheduling policies to minimize the average AoI.
A simple probabilistic correlation model is considered in \cite{tripathi2024Ton}, where each sensor can capture information about other sensors with a specific probability following a Bernoulli distribution, based on which \cite{tripathi2024Ton} proposed scheduling policies with performance guarantees.
Additionally, \cite{Tong2022TWC} addressed a multichannel scheduling problem in the correlated network and provides a scheduling policy within the Multi-Armed Bandit framework. \cite{Ramakanth2024TMC} jointly optimized AoI and estimation error under correlated sources and designed low-complexity scheduling policies. 
The above works focus mainly on AoI optimization under centralized networks. Recently, \cite{RAcorrelated} optimized AoI in random access networks under the correlated model proposed in \cite{tripathi2024Ton} using a Sequential Quadratic Programming (SQP). 
However, it did not explore how a sensor's access behavior influences competition within the correlated network.

\subsection{Timely and Energy-Efficient Information Delivery} \label{subsec: EE in RA}
Existing studies reviewed in Section \ref{subsec: AoI in RA} and Section \ref{subsec: AoI in correlated} neglect the impact of finite energy constraints while investigating optimal AoI in wireless networks. This omission may result in excessive energy consumption and a reduced lifetime of MTDs, and the optimal system configuration can differ substantially in EA-IoT.
Several works \cite{zhq_EE, Anshan2024EE, 24WangXliotj, fangming2022wcnc, fangming2022TGCN, yanbo2022Globecom, Jung2023Access} have instead focused on the energy efficiency of random access networks. In particular, \cite{zhq_EE} and \cite{Anshan2024EE} derived explicit expressions for device lifetime, obtained the corresponding optimal energy efficiency, and conducted a comprehensive analysis in Aloha-type M2M networks. Building on this line, \cite{24WangXliotj} developed an analytical framework for on-demand-sleep-based slotted Aloha, enabling lifetime analysis and optimization by introducing a dynamic active timer for devices with empty buffers.
The energy efficiency of ad-hoc scenarios, where multiple transmitter–receiver pairs coexist and compete for the channel, was analyzed in \cite{fangming2022wcnc, fangming2022TGCN}. Moreover, \cite{yanbo2022Globecom} and \cite{Jung2023Access} investigated the energy efficiency of Wi-Fi networks under throughput constraints.


Prior studies have typically explored AoI (both uncorrelated, as discussed in Section \ref{subsec: AoI in RA}, and correlated, in Section \ref{subsec: AoI in correlated}) and energy efficiency (i.e., \cite{zhq_EE, Anshan2024EE, 24WangXliotj, fangming2022wcnc, fangming2022TGCN, yanbo2022Globecom, Jung2023Access}) in isolation, focusing exclusively on optimizing a single performance metric. However, in networks with finite energy budgets and sensor correlations, new challenges arise, as the optimal strategy must jointly balance the tradeoff between AoI and energy efficiency across different correlation structures.
Consequently, in EA-IoT, the problem of leveraging spatial correlations among sensors to analyze the AoI of individual nodes, while simultaneously ensuring energy efficiency and mitigating severe contention, remains largely unexplored.
Addressing this challenge requires (i) understanding the spatial relationships among sensors, (ii) establishing analytical models that jointly characterize AoI and energy efficiency under correlation, and (iii) optimizing transmission strategies tailored to different correlation structures that balance energy-efficient, timely updates with minimal contention.

In this paper, we study a heterogeneous correlated random access network where each sensor adopts a distinct transmission probability, and both the AoI and energy efficiency of a given sensor are affected by other sensors under various correlation structures. 
Our contributions are summarized as follows:
\begin{itemize}
    \item[1)]  
    We develop an analytical model to characterize both AoI and energy efficiency in correlated random access networks. Specifically, we model the AoI evolution of each sensor as a discrete-time Markov process and derive the steady-state probability distribution of the corresponding Markov chain. Based on this distribution, we obtain an explicit expression for the long-term average AoI of each sensor. Furthermore, by accounting for different power consumption levels in the transmission and idle states of each sensor, we derive a closed-form expression for energy efficiency.
    \item[2)] By constraining sensors to adopt the same transmission probability, we develop three transmission strategies: the age-optimal strategy (Lemma \ref{lemma 3}), the energy-efficiency optimal strategy (Lemma \ref{lemma 4}), and the Pareto-optimal strategy (Theorem \ref{theorem 1}), which jointly optimizes AoI and energy efficiency. The corresponding transmission probabilities are denoted as $q_A^*$, $q_E^*$, and $q^*$, respectively. To determine $q^*$, we further propose a bounded exhaustive search method that guarantees both high reliability and low computational complexity. The complexity of this method is $\mathcal{O}(\frac{q_A^*-q_E^*}{q_{\epsilon}})$, which is bounded by $\mathcal{O}(\frac{1}{nq_{\epsilon}})$, and this bound decreases as the network size increases. Additionally, our theoretical analysis demonstrates that there exists a correlation gain, which substantially improves both AoI and energy efficiency.
    \item[3)] 
    Note that a sensor should refrain from transmitting when the marginal improvement in AoI or energy efficiency from its own successful transmission is smaller than the corresponding gains obtained through correlated neighbors’ transmissions. This observation highlights the need for carefully tuned transmission strategies. To this end, we propose the Multi-Start Projected Adaptive Moment Estimation (MS-PAdam) method to optimize network performance and evaluate the impact of the resulting optimal transmission strategy. The proposed algorithm converges efficiently to the optimal AoI and energy-efficiency tradeoff while identifying the corresponding transmission probabilities. Our results demonstrate that sensor competition can be substantially alleviated in heterogeneous correlated random access networks while maintaining optimal AoI and energy efficiency.
\end{itemize}

The remainder of this paper is organized as follows. Section \ref{System Model And Preliminary Analysis} establishes an analytical framework to characterize the AoI dynamics in correlated random access networks. Based on the proposed framework, Section \ref{Average AoI Analysis And Optimization Problem Formulation} derives closed-form expressions for the long-term average AoI and energy efficiency of each sensor and formulates the optimization problem.
Section \ref{homogeneousAoI section} derives three explicit transmission strategies under the constraint that all sensors adopt the same transmission probability, and further proposes a bounded exhaustive search method with provable complexity guarantees.
The global optimization in heterogeneous networks is carried out in Section \ref{Global AoI section}. Finally, concluding remarks are present in Section \ref{Conclusion}.

\section{System Model And Preliminary Analysis} \label{System Model And Preliminary Analysis}
\begin{figure*}[t]
\vspace{-0.5cm}
\centering
\subfloat[]{\includegraphics[width=3.6in,height=1.5in]{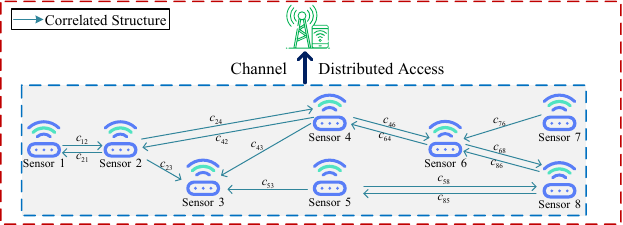}
	\label{fig: network illustrate}}\quad\quad
\subfloat[]{\includegraphics[width=2.9in,height=1.5in]{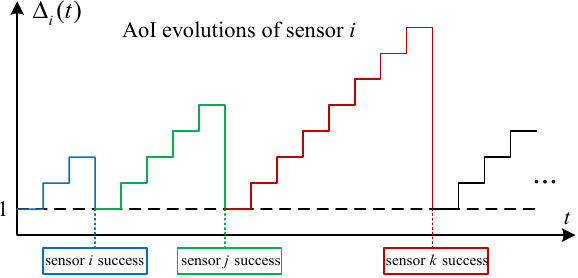}
        \label{fig: aoi_evolutions}}
    \caption{(a) Network illustration for correlated sensors.  (b) AoI evolutions of the sensor $i$.}
    \label{fig: system_model}
\end{figure*}
\subsection{Random Access Network Characterization}
As shown in Fig. \ref{fig: network illustrate}, we consider a status update system consisting of $n$ heterogeneous sensors and one base station, where $n$ sensors observe part of physical processes from the environment, and the base station will play a role in utilizing the transmitted information to estimate the status of the remote system.
Generally, we consider using the typical random access protocol, i.e., Slotted Aloha, in M2M communications system.
In heterogeneous slotted Aloha, the sensor $i$, for $i \in \mathcal{N}=\{1,2,\ldots,n\}$, transmits a status update with probability $q_i$. When the transmission fails, the sensor regenerates a new status update with probability $q_i$ in the next time slot. In this paper, we adopt the classic collision model in the receiver, i.e., a status update is successfully delivered if and only if there is no concurrent packet transmission in the channel.
We also employ the \textit{generate-at-will} strategy to manage the status update packets of each sensor, i.e., once a sensor decides to send its status update to the receiver, it generates a fresh status update before the transmission. If the transmission fails, the current status update will be discarded. When a failed sensor retransmits, it generates a new status update. For simplicity, let $a_i(t)$ be an indicator variable that denotes whether sensor $i$ successfully transmits the status update to the base station in time slot $t$. According to the distributed nature of the access behavior of each slotted Aloha sensor, we have $\Pr \{ a_i(t)=1\}= q_ip_i(t) $, and $\Pr \{ a_i(t)=0\}= 1-q_ip_i(t) $, where $p_i(t)$ denotes the successful transmission probability of sensor $i \in \mathcal{N}$ at time slot $t$. 
\begin{remark}
    Under centralized scheduling strategies, e.g., \cite{tripathi2024Ton}, the base station examines the AoI values of all nodes and selects one to query for an update. Although it is well known that centralized scheduling in \cite{tripathi2024Ton} achieves better performance, this process relies on polling mechanisms, which may be unsupported at the Medium Access Control (MAC) layer and can incur significant overhead in large-scale networks. These limitations highlight the importance of exploring distributed approaches to maintain information freshness in wireless settings.
\end{remark}

\subsection{Age of Information in Correlated Networks: Probabilistic Correlation Modeling}
We leverage AoI to evaluate the information freshness of the sensors at the base station. AoI measures the time elapsed since the latest received packet at the base station is generated from the sensor, where it grows linearly with time in the absence of new updates at the destination and reduces to the time elapsed since the generation of the delivered status update upon receiving a new status update. A pictorial example of the AoI evolution of the sensor $i$ is given in Fig. \ref{fig: aoi_evolutions}. Formally, this process can be expressed as
\begin{equation}
    \Delta_i(t) = t-G_i(t),
\end{equation}
where $G_i(t)$ is the generation time of the latest status update transmitted by sensor $i$ successfully received at the base station at time $t$. It should be noted that even the successful transmission of other sensors, e.g., sensor $j$ and sensor $k$ in Fig. \ref{fig: aoi_evolutions}, can reduce the AoI of sensor $i$.

We characterize the correlation structure between sensors using a matrix $\mathbf{C} \in \mathbb{R}^{n \times n}$. At the beginning of each time slot, sensor $i$ samples information about its own state and transmit to the receiver with probability (w.p.) $q_i$. Additionally, the status update captured by sensor $i$ also includes information about the current state of sensor $j$ w.p. $c_{ij} \in [0,1]$, which can occur due to overlapping fields of view between sensors $i$ and $j$ or spatial correlation in the monitored processes. We assume this information sharing or overlap occurs independently for each ordered sensor pair and across time. Consequently, a value of $c_{ij}$ = 0 indicates that sensor $i$ never captures information about sensor $j$, while $c_{ij}=1$ suggests that sensor $i$ has complete details on sensor $j$ at all times. Therefore, the network correlation structure can be described by a matrix $\textbf{C}$, \textit{where each off-diagonal element $c_{ij}$ represents the degree of correlation between sensors}. When $c_{ii} = 1$ for all sensors in $\mathcal{N}$, each sensor is assumed to always have information about itself. The model also accommodates cases where a sensor intermittently lacks self-information by setting $c_{ii} < 1$. 

Fig. \ref{fig: network illustrate} illustrates an example of a correlated random access network, where sensors capture partial status information from their neighbors and transmit updates to a base station.  
When sensor $i$ successfully transmits to the base station, it conveys its own state w.p. $c_{ii}$ and shares correlated information from other sensors w.p. $c_{ij}$ for $j \in [1,n]$. Consequently, updates received at the base station are often correlated, encompassing data from multiple correlated sensors. For instance, in Fig. \ref{fig: network illustrate}, when sensor 2 successfully transmits, its update includes its own status w.p. $c_{22}$ and, w.p. $c_{21}$, $c_{23}$, and $c_{24}$, may also incorporate information from sensors 1, 3, and 4, respectively.
Based on the correlation structure described above, the complete AoI evolution of sensor $i$ can be summarized as illustrated in Fig. \ref{fig: aoi_evolutions}. Both successful transmissions by sensor $i$ and by any other sensor correlated with sensor $i$, e.g., sensor $j$, $k$ in Fig. \ref{fig: aoi_evolutions}, influence the AoI evolution of sensor $i$ w.p. $c_{ii}$, $c_{ji}$, and $c_{ki}$, respectively.

Let $D_{ij}(t)$ be an indicator variable representing whether the current update at sensor $i$ also captures the information about sensor $j$ in slot $t$. Note that $D_{ij}(t) \sim Bern(c_{ij})$ across pairs $(i,j)$ and independently over time.
Given this correlation structure, represented by matrix $\textbf{C}$, the AoI for sensor $i$ at the base station evolves in the correlated network can be summarized as follows

\begin{equation}\label{AoI_definition1}
\Delta_i(t)=\!\!\left\{\!\!\!\!\!
\begin{array}{ll}
\,\,\,1&\text{if}\,\,\sum_{j\in\mathcal{N}} a_j(t)D_{ji}(t)=1 \\
\,\,\,\min\{\Delta_i(t-1)+1,\hat{\Delta}\} & \text{otherwise,}\,\,\,\,\,\,
\end{array}\right.
\end{equation}
where the parameter $\hat{\Delta}$ represents the maximum permissible AoI. An upper bound is imposed on AoI because excessively outdated packets are ineffective for time-sensitive applications. The specific value of $\hat{\Delta}$ is finite and determined by the requirements of the given application \cite{wang2022iotj}.
According to the description about the indicator variables $a_i(t)$ and $D_{ij}(t)$, the evolution of the AoI can be formulated as
\begin{equation}\label{AoI_definition2}
\Delta_i(t)=\left\{\!\!
\begin{array}{ll}
1 & \text{w.p.}\,\,p_{i}^{r}(t), \\
\min\{\Delta_i(t-1)+1,\hat{\Delta}\} & \text{w.p.}\,\,1-p_{i}^{r}(t),
\end{array}\right.
\end{equation}
where $p_{i}^{r}$ denotes the AoI reset probability of sensor $i$.
Note that for any sensor $j$ successfully transmits, the AoI of the sensor $i \in \mathcal{N}$ will be reset w.p. $p_j(t) q_j c_{ji}$. Therefore, the AoI reset probability of sensor $i$ at time slot $t$ is given by
\begin{equation} \label{AoI_reset_prob}
    p_{i}^{r}(t) =  \sum_{j\in \mathcal{N}} p_j(t) q_j c_{ji}.
\end{equation}

\begin{remark}
There are several motivating applications for our correlated monitoring framework. One prominent example arises in wireless sensor networks where updates collected by sensors exhibit spatial correlation. If the underlying spatial process and sensor locations remain relatively stable, the correlation structure $\mathbf{C}$ among sensors can be considered time-invariant. 
Another relevant scenario involves multiple static intelligent monitors coordinating with neighboring sensors to monitor an unoccupied area (e.g., an unoccupied room, a smart factory, etc.), while a remote central data fusion center aims to receive timely updates for situational awareness. This setting closely aligns with our model, as the positions of the intelligent monitors within the area remain fixed.
\end{remark}

\subsection{Energy Consumption Model}
Let $P_I$ and $P_T$ denote the basic power consumption in the idle and transmission states, respectively, as determined by the practical network configuration.
Assume a finite amount of energy for each sensor $E$, which is normalized to the time slot length. 
Due to limited energy, each sensor can only successfully deliver a limited number of packets during its lifetime. As such, a high energy efficiency is required. To quantify energy efficiency, we focus on the lifetime throughput metric \cite{Anshan2024EE}, $U_i$, defined as \textit{the average number of packets sensor $i$ can successfully deliver during its lifetime with a given initial energy $E$}.
As a result, the energy efficiency of each sensor $\xi_i$, which is defined as \textit{the ratio of the lifetime throughput $U_i$ to energy budget $E$}, i.e., $\xi_i = \frac{U_i}{E}$. 

In the following section, both the closed-form expressions of the AoI and energy efficiency will be characterized.

\section{AoI and Energy Efficiency: Expression Derivation and Optimization Problem Setup} \label{Average AoI Analysis And Optimization Problem Formulation}

\subsection{Average Correlated AoI Analysis}
\begin{figure}[t]
\centering
\includegraphics[width=0.5\textwidth]{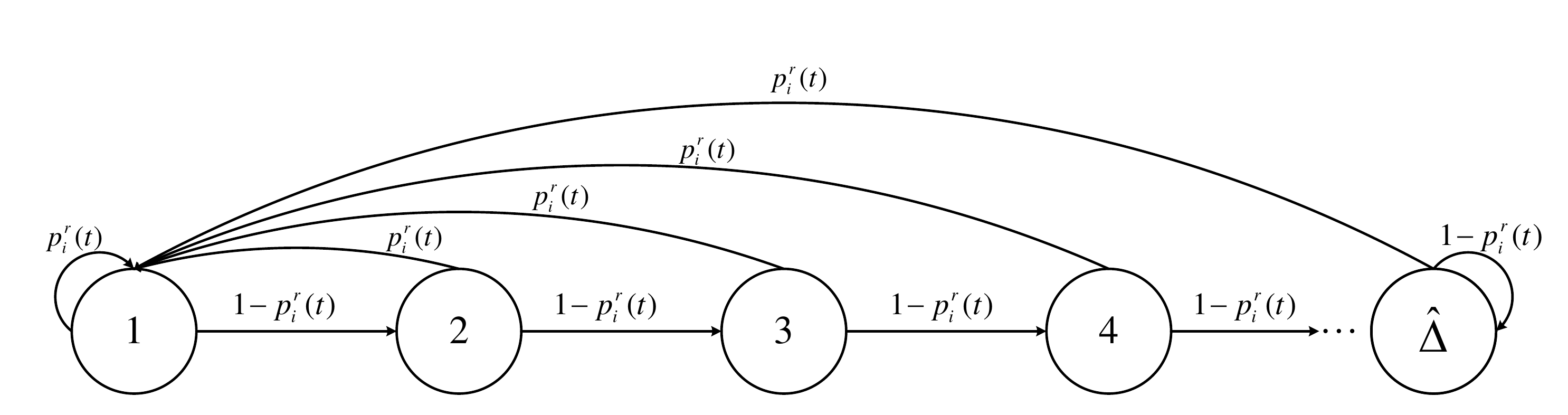}
\caption{AoI dynamics process of sensor $i$ in the correlated random access network.}
\label{AoI_Markovdynamics}
\end{figure}

According to the AoI dynamics process given in \eqref{AoI_definition2}, we know that the AoI dynamics of each sensor can be regarded as a Discrete-Time Markov Process under a discrete-time Markov chain. The Markov chain of each sensor is illustrated in Fig. \ref{AoI_Markovdynamics}, where the AoI of sensor $i$ resets upon the successful transmission of itself or other correlated sensors; otherwise, it increases by one.

It can be easily shown that the Markov chain in Fig. \ref{AoI_Markovdynamics} is uniformly strongly ergodic if and only if the limit
\begin{equation}
    \lim_{t\rightarrow \infty} p_{i}^{r}(t) = p_{i}^{r}
\end{equation}
exists \cite{iosifescu2014finite}, which is equivalent to 
\begin{equation}
    \lim_{t\rightarrow \infty} p_{i}(t) = p_{i}.
\end{equation}
Under the collision model, a transmission from sensor $i$ is successful if and only if none of the other $n-1$ sensors access the channel. Therefore, $p_i$ and $p_{i}^{r}$ can be written as 
\begin{equation}\label{p_success and AoI_reset_prob}
    p_i =\prod_{k\in \mathcal{N}\setminus i}(1-q_k);~~ p_{i}^{r} = \sum_{j\in \mathcal{N}} p_j q_j c_{ji}.
\end{equation}
The following Lemma presents the steady-state probability distribution of the Markov chain in Fig. \ref{AoI_Markovdynamics} and the expression of the long-term average AoI of sensor $i$:
\begin{lemma} \label{Lemma 1}
    The steady-state probability distribution of the Markov chain in Fig. \ref{AoI_Markovdynamics} can be obtained as 
\begin{equation}
\pi_{i}^{k} = 
    \begin{cases}
        (1-p_{i}^{r})^{k-1}p_{i}^{r} & k=1,2,\ldots,\hat{\Delta}-1\\
        (1-p_{i}^{r})^{k-1} & k=\hat{\Delta}\\
\end{cases}.
\end{equation}
where $p_{i}^{r}$ is given by \eqref{p_success and AoI_reset_prob}, and the long-term average AoI of sensor $i$ is given by
\begin{equation}\label{average AoI for one node}
\begin{aligned}
       \bar{\Delta}_i &= \frac{1-(1-p_{i}^{r})^{\hat{\Delta}}}{p_{i}^{r}}    \\   &=\frac{1-\left(1- \sum_{j=1}^n \prod_{k\in \mathcal{N}\setminus i}(1-q_k) q_j c_{ji}\right)^{\hat{\Delta}}}{\sum_{j=1}^n \prod_{k\in \mathcal{N}\setminus i}(1-q_k) q_j c_{ji}}.
\end{aligned}
\end{equation}
\end{lemma}
\begin{IEEEproof}
    Please see Appendix \ref{proof of lemma 1}.
\end{IEEEproof}

Finally, the network average AoI, $ \bar{\Delta}$, defined as the sum of the AoI values of all nodes, is expressed as follows
\begin{equation}\label{network AoI}
        \bar{\Delta}=\sum_{i=1}^{n}\bar{\Delta}_i = \sum_{i=1}^{n}\frac{1-\left(1- \sum_{j=1}^n \prod_{k\in \mathcal{N}\setminus i}(1-q_k) q_j c_{ji}\right)^{\hat{\Delta}}}{\sum_{j=1}^n \prod_{k\in \mathcal{N}\setminus i}(1-q_k) q_j c_{ji}}.
\end{equation}

\subsection{Energy Efficiency Analysis}
The following Lemma presents the energy efficiency of each sensor $i$ for $i\in \mathcal{N}$.
\begin{lemma} \label{Lemma 2}
    The lifetime throughput of sensor $i$ is given by
\begin{equation}\label{eq: lifetimethp}
    U_i = \frac{E\sum_{j\in \mathcal{N}} p_j q_j c_{ji}}{P_I+q_i(P_T-P_I)},
\end{equation}
and the energy efficiency of sensor $i$ is given by
\begin{equation}\label{eq: EE}
    \xi_i = \frac{\sum_{j\in \mathcal{N}} p_j q_j c_{ji}}{P_I+q_i(P_T-P_I)}.
\end{equation}
\end{lemma}
\begin{IEEEproof}
    Please see Appendix \ref{proof of lemma 2}.
\end{IEEEproof}
Therefore, the aggregate network lifetime throughput $U$ and energy efficiency $\xi$, defined as the sum of the lifetime throughput and energy efficiency of all nodes, can be expressed as $U = \sum_{i\in \mathcal{N}} U_i, \xi = \sum_{i\in \mathcal{N}} \xi_i,$, respectively.

\subsection{Problem Formulation}
In practical networks, it is generally desirable to maximize the energy efficiency of each sensor while minimizing the AoI performance. Therefore, our objective is to achieve this dual goal by optimizing the transmission probabilities.
The optimization problem can be defined as
\begin{equation}\label{eq: Problem_formulate}
    \mathcal{P}_{0}\left\{
\begin{array}{ll}
    \min \limits_{\boldsymbol{q}} & \gamma_{1}\bar{\Delta} - \gamma_{2}\xi \\
    \text{s.t.} & \boldsymbol{q} = \{q_1, q_2,...,q_n\} \in [0,1]^n ,
\end{array}
\right.
\end{equation}
where $\boldsymbol{q} = (q_1,\ldots,q_n)$ denotes the transmission probability vector, and $\gamma_{1}$ and $\gamma_{2}$ are scaling factors introduced to balance the magnitudes of the two optimization objectives. Specifically, $\gamma_{1}$ is dimensionless, whereas $\gamma_{2}$ has units of joules per packet to ensure dimensional consistency in problem \eqref{eq: Problem_formulate}.

\section{Homogeneous AoI and Energy Efficiency Optimization}\label{homogeneousAoI section}

We begin by optimizing AoI and energy efficiency in the homogeneous scenario, which can be viewed as a special case of the general optimization problem \eqref{eq: Problem_formulate}, thereby offering a comprehensive understanding of the resulting transmission strategies. 
In this case, the transmission probability of each sensor is subjected to $q_i = q$, for $i\in\mathcal{N}$, and the successful transmission probability $p_i=p=(1-q)^{n-1} {\overset{n\gg 1}{\approx}} \exp(-nq)$. 
Therefore, the optimization problem \eqref{eq: Problem_formulate} can be rewritten as
\begin{equation}\label{eq: Problem_formulate_homo}
    \mathcal{P}_{1}\left\{
\begin{array}{ll}
    \min \limits_{\boldsymbol{q}} & \gamma_{1}\bar{\Delta} - \gamma_{2}\xi \\
    \text{s.t. } & \boldsymbol{q} = \{q_1, q_2,...,q_n\} \in [0,1]^n \\
    & q_i = q_j \quad \forall{i,j \in \mathcal{N}.}
\end{array}
\right.
\end{equation}
\subsection{Main Analytical Results}
The following Lemmas present the optimal network average AoI $\bar{\Delta}^*$ and the optimal network energy efficiency $\xi^*$ for the homogeneous correlated network, along with their corresponding optimal transmission probabilities. 

\begin{lemma}\label{lemma 3}
    The optimal network average AoI is given by
    \begin{equation}
        \bar{\Delta}^* =\sum_{i=1}^{n} \frac{n-n\exp \left\{-{\hat{\Delta}\exp(-1)\sum_{j=1}^n c_{ji}/n} \right\}}{\exp(-1)\sum_{j=1}^n c_{ji}}.
    \end{equation}
    The optimal AoI reset probability of each sensor $i$ and the optimal transmission probability that minimizes network AoI $\bar{\Delta}$ are given by
    \begin{equation} \label{eq: homoAoI_Opt}
        p_{i}^{r*} = \frac{\exp(-1)\sum_{j=1}^n c_{ji}}{n}, q_A^*=\frac{1}{n}.
    \end{equation}
\end{lemma}
\begin{IEEEproof}
    Please see Appendix \ref{proof of lemma 3}.
\end{IEEEproof}

From Lemma \ref{lemma 3}, we conclude that although the optimal transmission probability of all sensors is still $q_i^* = \frac{1}{n}$ in the homogeneous network, both the optimal AoI reset probability $p_{i}^{r*}$ and network average AoI $\bar{\Delta}^*$ are improved by the correlated factor $\sum_{j=1}^n c_{ji}$.
As a result, Lemma \ref{lemma 3} explicitly demonstrates that the spatial correlation of sensors can enhance information freshness in random access networks.

\begin{lemma}\label{lemma 4}
    The optimal network energy efficiency $\xi^*$ is given by
    \begin{equation}
        \xi^* = \sum_{i\in \mathcal{N}}\frac{\sum_{j\in \mathcal{N}}  c_{ji}}{\frac{P_I}{q_E^*\exp(-nq_E^*)}+\frac{P_T-P_I}{\exp(-nq_E^*)}}.
    \end{equation}
    The optimal transmission probability for sensor $i$ that maximizes $\xi^*$ is given by
    \begin{equation}\label{eq: homoEE_opt}
         q_E^*=\frac{\sqrt{1+\frac{4}{n}\left(\frac{P_T}{P_I}-1\right)}-1}{2\left(\frac{P_T}{P_I}-1\right)}.
    \end{equation}
\end{lemma}
\begin{IEEEproof}
    Please see Appendix \ref{proof of lemma 4}.
\end{IEEEproof}

\begin{theorem}\label{theorem 1}
    By combining Lemmas \ref{lemma 3} and \ref{lemma 4}. we derive that the optimal transmission probability $q^*$ in homogeneous networks lies within the interval $q^*\in [q_E^*, q_A^*]$.
\end{theorem}
\begin{IEEEproof}
    Please see Appendix \ref{proof of theorem 1}.
\end{IEEEproof}
Theorem \ref{theorem 1} establishes that the optimal transmission probability $q^*$, which solves the constrained optimization problem \eqref{eq: Problem_formulate_homo}, lies within the interval $[q_E^*, q_A^*]$. Note that $q_E^* \leq q_A^*=\frac{1}{n} \ll 1$, where $n$ is the number of sensors in the network. This implies that only a narrow range of $q$ needs to be evaluated. 
Consequently, we employ an exhaustive search method (i.e., Algorithm \ref{alg: homoExhaustive}), both highly reliable and computationally efficient. The complexity is $\mathcal{O}(\frac{q_A^*-q_E^*}{q_{\epsilon}}) < \mathcal{O}(\frac{1}{nq_{\epsilon}})$, significantly reduced due to the confined search interval $[q_E^*, q_A^*]$ proposed in Theorem \ref{theorem 1}.

\begin{algorithm}[t]
\caption{Bounded exhaustive search method for homogeneous networks optimization}
\label{alg: homoExhaustive}
\begin{algorithmic}[1]
\Require Search precision $q_{\epsilon}$, search interval bounds $q_E^*$, $q_A^*$, and scaling factor $\gamma_{1},\gamma_{2}$
\State Initialize: Optimal solution $q^* \gets \emptyset$, minimum AoI $\bar{\Delta}^* \gets +\infty$, and maximum energy efficiency $\xi^* \gets -\infty$
\State Generate candidate set $\mathcal{Q} \gets \{q_1,...,q_N\}$ by discretizing $[q_E^*, q_A^*]$ with step size $q_{\epsilon}$
\For{each candidate $q \in \mathcal{Q}$}
    \State Compute network Age of Information $\bar{\Delta}(q)$ via \eqref{network AoI}
    \State Compute energy efficiency $\xi(q)$ via \eqref{eq: EE}
    \State Evaluate objective function $J(q) \gets \gamma_{1}\bar{\Delta}(q) - \gamma_{2}\xi(q)$
    \If{$J(q) < \gamma_{1}\bar{\Delta}^* - \gamma_{2}\xi^*$}
        \State Update optimal solution: $q^* \gets q$, $\bar{\Delta}^* \gets \bar{\Delta}(q)$, $\xi^* \gets \xi(q)$
    \EndIf
\EndFor
\Return $(q^*, \bar{\Delta}^*, \xi^*)$
\end{algorithmic}
\end{algorithm}

\begin{figure*}[t]
\vspace{-1cm}
\centering
\subfloat[]{\includegraphics[width=3.1in,height=2.1in]{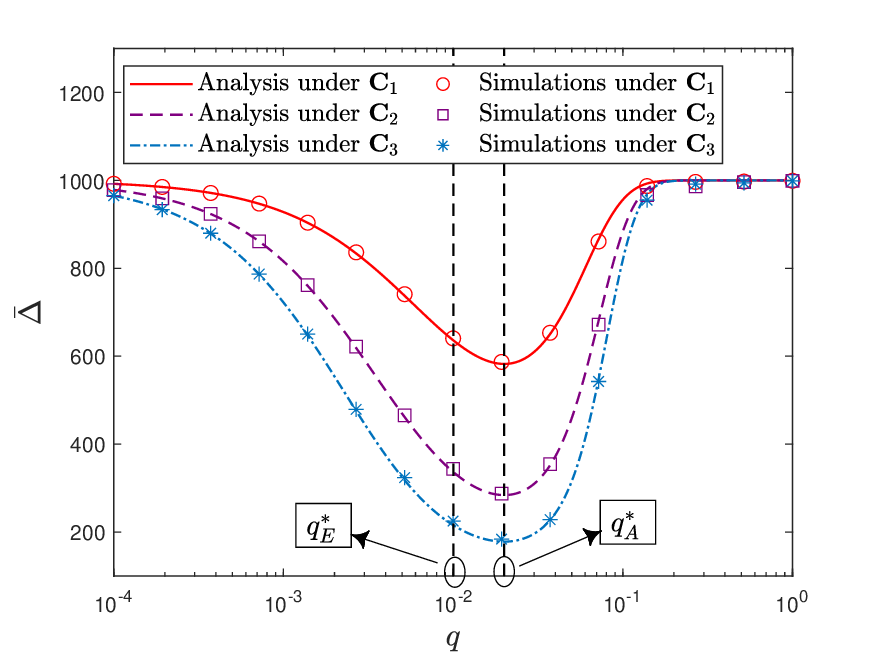}
	\label{fig: homo_AoIsim}}\quad\quad
\subfloat[]{\includegraphics[width=3.1in,height=2.1in]{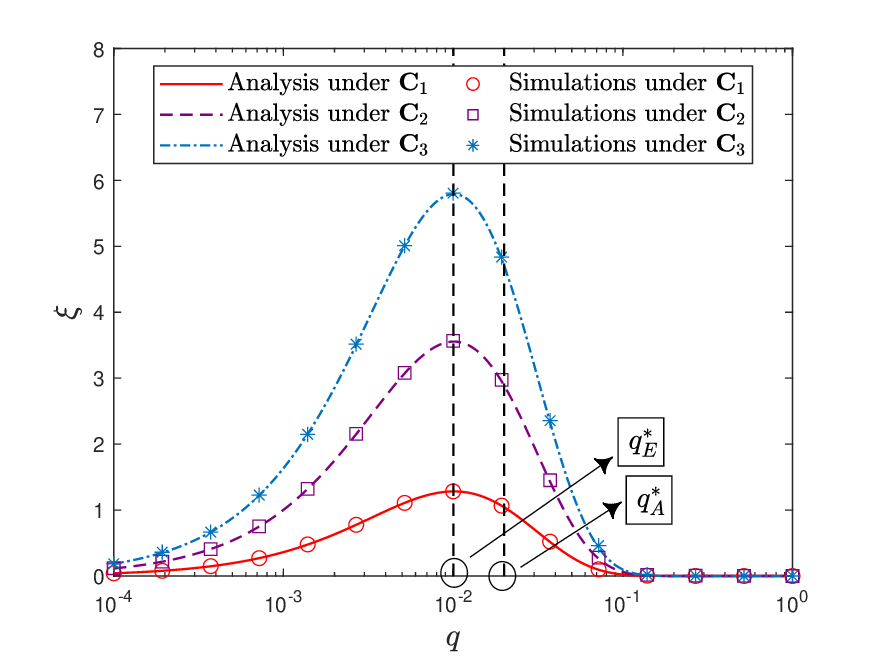}
        \label{fig: homoEEsim}}
    \caption{(a) Network average AoI $\bar{\Delta}$ and (b) energy efficiency $\xi$ versus transmission probability $q$ in the homogeneous network under different correlation matrix $\textbf{C}$ settings. The off-diagonal elements $c_{ij}$ (i.e., degree of correlation) are randomly selected from $[0,0.3]$ for $\textbf{C}_1$, $[0.3,0.6]$ for $\textbf{C}_2$, and $[0.6,0.9]$ for $\textbf{C}_3$. Other parameters: $n=50, \hat{\Delta}=20, E = 1\times10^6, P_T = 100, P_I=1$.}
    \label{fig: homosim}
\end{figure*}

\subsection{Numerical Result in Homogeneous Network} \label{Homo_numerical_result}
To demonstrate the effect of correlation structure and validate Lemmas \ref{Lemma 1}-\ref{lemma 4}, we employ three distinct correlation matrices $\mathbf{C}$ (denoted as $\textbf{C}_1$, $\textbf{C}_2$, and $\textbf{C}_3$, respectively) in our simulations, as illustrated in Fig. \ref{fig: homosim}. The matrix $\textbf{C}_1$ denotes the low-correlation matrix, with diagonal elements $c_{ii} = 1$ (i.e., all the sensors have perfect knowledge of themselves) and off-diagonal elements $c_{ij}$ (i.e., degree of correlation) are randomly selected from $[0,0.3]$ for $\textbf{C}_1$. The matrices $\textbf{C}_2$ and $\textbf{C}_3$, referred to as the medium-correlation and high-correlation matrices, respectively, share the same structure as $\textbf{C}_1$, but the degree of correlation are randomly selected from $[0.3,0.6]$ and $[0.6,0.9]$, respectively.
Additionally, the simulations in Fig. \ref{fig: homosim} assume a network consisting of $n=50$ sensors. In each time slot, each sensor $i \in \mathcal{N}$ samples and transmits its status update with probability $q$ and transmit power $P_T=100$; otherwise, it remains idle with probability $1-q$ and idle power $P_I=1$. The maximum permissible AoI for each sensor is $\hat{\Delta}=20$, and the initial battery capacity is $E = 1 \times 10^6$. During the simulation, the AoI dynamics of each sensor follow \eqref{AoI_definition1}, and we compute the successfully delivered status updates to determine the lifetime throughput $U_i$ for each sensor $i \in \mathcal{N}$, from which the energy efficiency $\xi_i$ is computed. Finally, the theoretical network average AoI $\bar{\Delta}$ and energy efficiency $\xi$ are calculated using \eqref{network AoI} and \eqref{eq: EE}, respectively.

\begin{figure}[t]
\vspace{-0.5cm}
    \centering
    \includegraphics[width=0.5\textwidth]{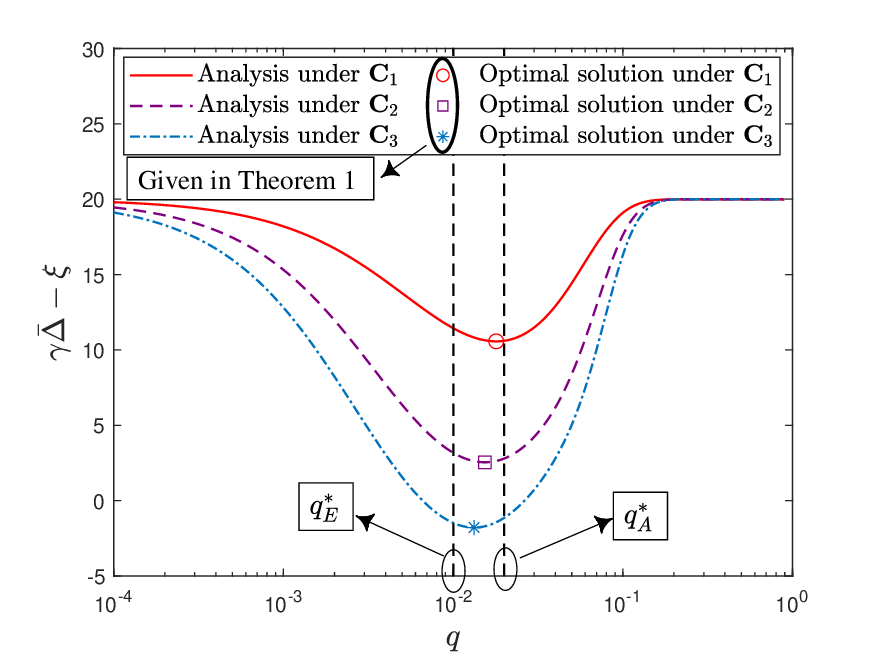}
    \caption{Object function in problem \eqref{eq: Problem_formulate_homo} versus $q$ and the corresponding optimal $q^*$ under different correlation matrix $\textbf{C}$ settings. $\textbf{C}_1$, $\textbf{C}_2$, $\textbf{C}_3$, and other parameters are the same as Fig. \ref{fig: homosim}. $\gamma_{1} = 0.02$, $\gamma_{2}=1$.}
    \label{fig: homoopt_validate}
\end{figure}

Both theoretical and simulation results in Fig. \ref{fig: homo_AoIsim} illustrate that the network average AoI $\bar{\Delta}$ monotonically decreases when $q<q_A^*$ and monotonically increases when $q>q_A^*$, the optimal network average AoI $\bar{\Delta}^*$ therefore is achieved at $q=q_A^*=\frac{1}{n}$. Similarly, Fig. \ref{fig: homoEEsim} illustrates that the optimal network energy efficiency $\xi^*$ occurs at $q=q_E^*$, as derived in Lemma \ref{lemma 3} and Lemma \ref{lemma 4}.
Furthermore, Fig. \ref{fig: homo_AoIsim} and Fig. \ref{fig: homoEEsim} clearly demonstrate the inherent trade-off between network average AoI $\bar{\Delta}$ and energy efficiency $\xi$ in homogeneous random access networks, where simultaneous optimization of both metrics is unachievable. This fundamental trade-off necessitates careful selection of the transmission probability. As established in Theorem \ref{theorem 1}, the optimal transmission probability $q^*$ lies within the interval $[q_E^*, q_A^*]$, requiring further evaluation to balance these competing objectives.
Based on Theorem \ref{theorem 1}, Fig. \ref{fig: homoopt_validate} demonstrates that the bounded exhaustive search method (i.e., Algorithm \ref{alg: homoExhaustive}) effectively identifies the optimal transmission probability $q^*$ that balances the fundamental trade-off between AoI performance and network energy efficiency.

\begin{remark}
    Since a central base station is available for managing all the sensors, the optimal configuration can be easily implemented. Specifically, the central base station calculates the optimal transmission probability according to Theorem \ref{theorem 1} in the homogeneous networks. Although computing $q_E^*$ and $q_A^*$ requires global information such as the number of sensors within the network, the base station can collect this global information from each sensor based on the feedback from sensors and send the optimal configuration to the sensors.
\end{remark}

\section{Global Network AoI and Energy Efficiency Optimization}\label{Global AoI section}
The preceding analysis explicitly determines the optimal transmission probability for the homogeneous case, which represents a special instance of the general optimization problem \eqref{eq: Problem_formulate}. A key characteristic of homogeneous networks is that all nodes share identical, non-zero transmission probabilities.
However, interestingly, when a sensor exhibits strong correlation with other nodes, the monitored area's state can often be inferred from neighboring sensors' updates. This correlation effect implies that even with reduced transmission frequency (or complete transmission abstention), such sensors can maintain satisfactory AoI performance and energy efficiency. We refer to this phenomenon as correlation gain.
Consequently, a sensor should suppress transmissions when the marginal AoI/energy efficiency improvement from its own successful transmission is outweighed by the corresponding gains from correlated neighbors' transmissions.

We then investigate the network AoI and energy efficiency performance under heterogeneous scenario, where the access behaviors of all sensors are different.
We therefore consider solving the global network average AoI optimization problem proposed in \eqref{eq: Problem_formulate}.
We next analyze and optimize network average AoI and energy efficiency in heterogeneous scenarios, where sensors exhibit distinct access behaviors. Given the derived closed-form expressions for AoI and energy efficiency, we propose a gradient-based optimization framework to address distributed access behaviors within sensors, which solves the global optimization problem formulated in Eq. \eqref{eq: Problem_formulate}.
\subsection{Gradient-based Network Optimization Framework}
Since the closed-form expression of $\bar{\Delta}$ and $\xi$ has been derived, the gradient expression of the network average AoI with respect to (w.r.t.) the transmission probability vector $\boldsymbol{q}$, denoted as $\nabla ( \gamma_{1}\bar{\Delta}(\boldsymbol{q})-\gamma_{2}\xi(\boldsymbol{q}))$ can be directly computed. 
Note that both $\bar{\Delta}$ and $\xi$ are determined by the transmission probability vector $\boldsymbol{q} = \{q_1,...,q_n\}$ and the correlation structure $\textbf{C}$, the gradient of $\gamma_{1}\bar{\Delta}(\boldsymbol{q})-\gamma_{2}\xi(\boldsymbol{q})$ w.r.t. $\boldsymbol{q}$ is also a vector with $n$ dimensions. We first provide the expression for the partial derivative of $\bar{\Delta}$ and $\xi$ concerning any specific transmission probability $q_s$

    \begin{equation}\label{eq: diff_Delta_to_q}
        \frac{\partial \bar{\Delta}}{\partial q_s} =
             \sum_{i=1}^n \left[ \frac{\hat{\Delta} \left( 1 - p_{i}^{r} \right)^{\hat{\Delta} - 1} }{p_{i}^{r}} \frac{\partial  p_{i}^{r}}{\partial q_s} - \frac{\left( 1 - \left( 1 - p_{i}^{r} \right)^{\hat{\Delta}} \right) }{(p_{i}^{r})^2} \frac{\partial  p_{i}^{r}}{\partial q_s} \right],
    \end{equation}
where $s=1,2,\ldots,n$. $\frac{\partial  p_{i}^{r}}{\partial q_s}$ is given by
\begin{equation}\label{eq: diff_pir_qs}
\begin{aligned}
   \frac{\partial  p_{i}^{r}}{\partial q_s}  
   & = p_s c_{si} -\frac{\sum_{j=1,j\neq s}^n q_jp_j c_{ji}}{1-q_s},
\end{aligned}
\end{equation}
and
\begin{equation}\label{eq: diff_EE_to_q}
    \frac{\partial \xi}{\partial q_s} = \sum_{i=1}^n \partial\frac{ p_i^r/(P_I+q_i(P_T-P_I))}{\partial q_s}.
\end{equation}
Combining with \eqref{eq: diff_Delta_to_q}-\eqref{eq: diff_EE_to_q}, the gradient expression $\nabla ( \gamma_{1}\bar{\Delta}(\boldsymbol{q})-\gamma_{2}\xi(\boldsymbol{q}))$ can be derived.

Note that the gradient expression $\nabla ( \gamma_{1}\bar{\Delta}(\boldsymbol{q})-\gamma_{2}\xi(\boldsymbol{q}))$ exhibits: 1). $n$ consecutive multiplication terms of the transmission probability $q_i$ for $i\in \mathcal{N}$; 2). exponential dimensionality and complexity with the network size $|\mathcal{N}|$. These properties render the constrained global optimization problem in \eqref{eq: Problem_formulate} strongly non-convex. Consequently, obtaining a globally optimal closed-form solution $\boldsymbol{q}^*$ that effectively balances both network AoI and energy efficiency remains fundamentally challenging.

To overcome these challenges, we develop a hybrid approach combining multi-start \cite{marti2003multi} initialization with the Adam optimization method \cite{Kingma2014AdamAM} to solve the constrained, strongly non-convex problem in Eq. \eqref{eq: Problem_formulate}. This strategy maintains feasible iterates while effectively escaping poor local optima. Compared to standard Gradient Descent (GD), Adam's adaptive learning rate proves particularly advantageous for our high-dimensional optimization over the transmission probability vector $\boldsymbol{q}$, as it automatically adjusts step sizes for each parameter dimension.
To enhance global optimum discovery, we implement a diversified multi-start initialization strategy. Specifically, we generate $M$ initial vectors $\{\boldsymbol{q}_{0}^{(1)},\boldsymbol{q}_{0}^{(2)},\ldots,\boldsymbol{q}_{0}^{(M)}\}$ with constrained spatial distribution by enforcing a minimum pairwise Euclidean distance $d_{\min}$: 
\begin{equation}\label{eq: generate init points}
    \|\boldsymbol{q}^{(l)}-\boldsymbol{q}^{(r)}\|_2 \geq d_{\min}, \forall ~ l,r \in \{1,...,M\},
\end{equation}
This dispersion criterion enables broad solution space exploration without trapping in local optima. MS-PAdam leverages this initialization to efficiently probe promising convex regions, significantly boosting convergence reliability.

For each transmission probability vector, during each iteration, we calculate the current gradient $\nabla ( \gamma_{1}\bar{\Delta}(\boldsymbol{q}_t)-\gamma_{2}\xi(\boldsymbol{q}_t))$, upon which we can update the first-order moment vector $\boldsymbol{m}_t$ and the second-order moment vector $\boldsymbol{\upsilon}_t$ as follow \cite{Kingma2014AdamAM}:

\begin{small}
\begin{equation}\label{eq: Adamupdate}
\begin{cases}
        \boldsymbol{m}_t = \beta_1 \boldsymbol{m}_{t-1} + (1-\beta_1)\nabla ( \gamma_{1}\bar{\Delta}(\boldsymbol{q}_t)-\gamma_{2}\xi(\boldsymbol{q}_t)),\\
        \boldsymbol{\upsilon}_t = \max \left\{ \boldsymbol{\upsilon}_{t-1},\beta_2 \boldsymbol{\upsilon}_{t-1} + (1-\beta_2)\left(\nabla ( \gamma_{1}\bar{\Delta}(\boldsymbol{q}_t)-\gamma_{2}\xi(\boldsymbol{q}_t))\right)^2\right\},\\
        \hat{\boldsymbol{m}}_t = \frac{\boldsymbol{m}_t}{1-\beta_1^t},\\
        \hat{\boldsymbol{\upsilon}}_t = \frac{\boldsymbol{\upsilon}_t}{1-\beta_2^t},
\end{cases}
\end{equation}
\end{small}

\noindent where $\beta_1$ and $\beta_2$ are attenuation factors. In particular, we modify the update rule of the second-order moment vector $\boldsymbol{\upsilon}_t$ to ensure that the learning rate $\eta$ decreases monotonically, which in turn guarantees the convergence of Adam \cite{47409}. The transmission probability vector can be updated by:
\begin{equation}
    \tilde{\boldsymbol{q}}_t = \boldsymbol{q}_{t-1} - \frac{\eta \hat{\boldsymbol{m}}_t}{\sqrt{\hat{\boldsymbol{\upsilon}}_t}+\delta},
\end{equation}
where $\delta$ is a small positive constant to prevent division by zero errors.
When $\boldsymbol{q}_t$ has been updated, it should be projected onto the feasible region by using:
\begin{equation}
    \boldsymbol{q}_t = \min(1-\delta, \max(\delta,\tilde{\boldsymbol{q}}_t)).
\end{equation}
Finally, we check whether the new $\boldsymbol{q}$ is convergent using the following equation:
\begin{equation}\label{convergence condition}
    \|\boldsymbol{q}_{t}-\boldsymbol{q}_{t-1}\|_2 \leq \epsilon,
\end{equation}
where $\epsilon$ denotes difference tolerance. 
Detail Algorithm methods of the MS-PAdam approach are elaborated in Algorithm \ref{alg: MS-PAdam}.
According to Algorithm \ref{alg: MS-PAdam}, the optimal transmission probability vector $\boldsymbol{q}^*$ can be obtained to optimize the network’s average AoI and energy efficiency.
\begin{remark}
    In enterprise-scale distributed IoT networks with centralized base station control, optimal configuration deployment is readily achievable. Specifically, the central base station computes the optimal transmission probability using Algorithm \ref{alg: MS-PAdam}. The base station computes the optimal transmission probabilities using Algorithm \ref{alg: MS-PAdam}, which requires global network parameters including the number of sensors and the correlation structure $\mathbf{C}$. These parameters are acquired through sensor feedback channels during normal status updates. Following computation, the base station broadcasts the optimized configuration to all network nodes, enabling synchronous implementation.
\end{remark}

\begin{algorithm}[t]
\caption{Multi-Start Projected Adaptive Moment Estimation (MS-PAdam) method}
\label{alg: MS-PAdam}
\begin{algorithmic}[1]
\Require Initial learning rate $\eta$, tolerance $\epsilon$, attenuation factors $\beta_1$, $\beta_2$, small enough number $\delta$, scaling factor $\gamma_{1},\gamma_{2}$, and randomly generated $M$ transmission probability vectors $\{\boldsymbol{q}_{0}^{(1)},\boldsymbol{q}_{0}^{(2)},\ldots,\boldsymbol{q}_{0}^{(M)}\}$ under dispersion criterion in \eqref{eq: generate init points}.
\State Initialize $\boldsymbol{q}^* \leftarrow \emptyset$, $\bar{\Delta}^* \leftarrow \infty$
\For{$l = 1, 2, \ldots, M$} \Comment{Multi-start initialization}
    \State Initialize $\boldsymbol{q}_{0} \gets \boldsymbol{q}_{0}^{(l)}, \quad \boldsymbol{m}_0 \gets \emptyset, \quad \boldsymbol{v}_0 \gets \emptyset
    $
    \While{not converged} \Comment{Adam-based PGD}
    \State Compute the gradient based on \eqref{eq: diff_Delta_to_q}, \eqref{eq: diff_pir_qs} and \eqref{eq: diff_EE_to_q} 
    \State Update moment vector based on \eqref{eq: Adamupdate}
    \State Perform Adam update step
    \State Project onto feasible region $\boldsymbol{q}_t \in [0, 1]^n$
    \If{$\|\boldsymbol{q}_{t}-\boldsymbol{q}_{t-1}\|_2 \leq \epsilon$} \Comment{Check convergence} 
        \State \textbf{break}
    \EndIf
\EndWhile
\State Compute objective value $\bar{\Delta}^{(m)} = \bar{\Delta}(\boldsymbol{q}_{t})$
\If{$\bar{\Delta}^{(m)} < \bar{\Delta}^*$}
    \State Update optimal solution: $\boldsymbol{q}^* \leftarrow \boldsymbol{q}_{t}$, $\bar{\Delta}^* \leftarrow \bar{\Delta}^{(m)}$
\EndIf
\EndFor
\Return $\boldsymbol{q}^*$ and $\bar{\Delta}^*$
\end{algorithmic}
\end{algorithm}

\subsection{Complexity and convergence Analysis of MS-PAdam}

\begin{table}[t]
	\centering
	\caption{Parameter configuration of MS-PAdam.}
		\begin{tabular}{l|c}\toprule
			Parameters    & Values     \\ \midrule
            Small positive constant $\delta$            & $10^{-8}$            \\ 
            Start points $M$ & 10 \\
			Attenuation factors $\beta_1,\beta_2$              & 0.9,0.999    \\
			AoI upper limits of each sensor $\hat{\Delta}$            & 20            \\ 
			Initial learning rate $\eta$            & 0.005            \\ 
            Error tolerance $\epsilon$            & $10^{-4}$            \\
            Minimum Euclidean distance $d_{\min}$ & 1                   \\
            scaling factor $\gamma_{1},\gamma_{2}$ & 0.1,1\\
                \bottomrule
	\end{tabular}
	\label{table1}
\end{table}

In this subsection, we analyze the complexity and convergence of the proposed MS-PAdam algorithm. The MS-PAdam algorithm primarily consists of two nested loops. The outer loop (Multi-Start loop) runs $M$ times, each time initializing the optimization from a new random $\boldsymbol{q}$. The inner loop consists of Adam iterations for each starting point.
In each Adam iteration, we need to compute the gradient value $\nabla ( \gamma_{1}\bar{\Delta}(\boldsymbol{q})-\gamma_{2}\xi(\boldsymbol{q}))$, compute and update the first-moment estimate $\boldsymbol{m}_t$ and second-moment estimate $\boldsymbol{\upsilon}_t$, update the transmission probability vector $\tilde{\boldsymbol{q}}$, project the $\tilde{\boldsymbol{q}}$ onto the feasible region, and check the convergence of the algorithm. Consequently, the complexity of each Adam iteration is $\mathcal{O}(n)$. As a result, the overall computational complexity of the MS-PAdam algorithm is $\mathcal{O}(nM)$, where $M$ is the number of starting points, controlling the global search range, and $n$ is the number of sensors, determining the dimensionality of the optimization problem.

Subsequently, we conduct a detailed analysis of the convergence properties of MS-PAdam in this study. Based on the expression of $\bar{\Delta}$, $\xi$, and the gradient expression $\nabla ( \gamma_{1}\bar{\Delta}(\boldsymbol{q})-\gamma_{2}\xi(\boldsymbol{q}))$, it is evident that $\bar{\Delta}$ and $\xi$ are continuously differentiable w.r.t. $\boldsymbol{q}$. Furthermore, $\bar{\Delta}$ and $\xi$ are functions mapping satisfies $\mathbb{R}^n \rightarrow \mathbb{R}$ and, since each sensor's long-term average AoI and energy efficiency is bounded by $\hat{\Delta}$ and $1$, the network average AoI $\bar{\Delta}$ is consequently bounded below by $n\hat{\Delta}$ and $n$. Based on the modified Adam update equations \cite{47409} in \eqref{eq: Adamupdate} and the convergence analysis of the GD method in \cite{calamai1987projected}, we can conclude that
\begin{equation}\label{convergence_result}
    \lim_{t\to\infty} \| \boldsymbol{q}_{t} -\boldsymbol{q}_{t-1}\|_2 = 0.
\end{equation}
Moreover, the multi-start mechanism is an effective strategy in optimization algorithms for increasing the likelihood of convergence to the global optimum.
As a result, the MS-PAdam algorithm effectively converges to the optimal solution and consistently yields the optimal transmission probability vector $\boldsymbol{q}^*$. Below, we conducted the convergence performance comparison between the proposed MS-PAdam and conventional GD methods. The key presetting parameters are listed in Table \ref{table1}. We evaluate both the number of iterations required for the algorithm to converge from each initial condition and the performance of the objective function $\gamma_{1}\bar{\Delta}-\gamma_{2}\xi$ across different network sizes $n$. Subsequently, we compute the average number of iterations and the objective function's performance over $M$ initial conditions for each network size. As shown in Table \ref{convergence table}, for small-scale networks (e.g., $n=4$ to 8), both algorithms converge to the same performance. However, the proposed MS-PAdam algorithm achieves faster convergence and better performance as the network scale increases (e.g., $n=12$ to 16). As a result, MS-PAdam demonstrates favorable scalability and robustness for solving the promising transmission strategy in the heterogeneous random access network with correlated status updates.

\begin{table}[t]
	\centering
	\caption{The convergence performance of the proposed MS-PAdam algorithm and standard GD algorithm.}
	\resizebox{1\linewidth}{!}{
		\begin{tabular}{>{\centering}m{1.2cm}|>{\centering\arraybackslash}m{2cm}|>{\centering\arraybackslash}m{2cm}|>{\centering\arraybackslash}m{2cm}|>{\centering\arraybackslash}m{2cm}}
        \toprule
		Number of sensors    & Performance of GD & Performance of MS-PAdam & Average iterations of GD & Average iterations of MS-PAdam    \\
            \midrule
            $n=4$    & 2.525 & 2.541 & 292.3 & 399.5     \\
            $n=8$    & 6.744 & 6.744 & 144.4 & 252.7     \\
            $n=12$    & 13.216 & 10.506 & 535.1 & 268.4     \\
            $n=16$    & 27.595 & 17.455 & 928.9 & 464.2     \\
             \bottomrule
	\end{tabular}
}
	\label{convergence table}
\end{table}

\subsection{Numerical Results for MS-PAdam: Optimal Validation and Performance Analysis}
We proceed to evaluate the performance of the proposed MS-PAdam algorithm in determining optimal solutions across various correlated scenarios and various network sensors number.
Below we conduct a comprehensive performance comparison between the proposed MS-PAdam algorithm and four benchmark strategies: (1) random transmission strategy, (2) homogeneous age-optimal strategy (Lemma \ref{lemma 3}), (3) homogeneous energy-efficiency optimal strategy (Lemma \ref{lemma 4}), and (4) Pareto-optimal strategy (Theorem \ref{theorem 1}), which jointly optimizes AoI and energy efficiency. 

\begin{figure}[t]
    \centering
    \includegraphics[width=0.48\textwidth]{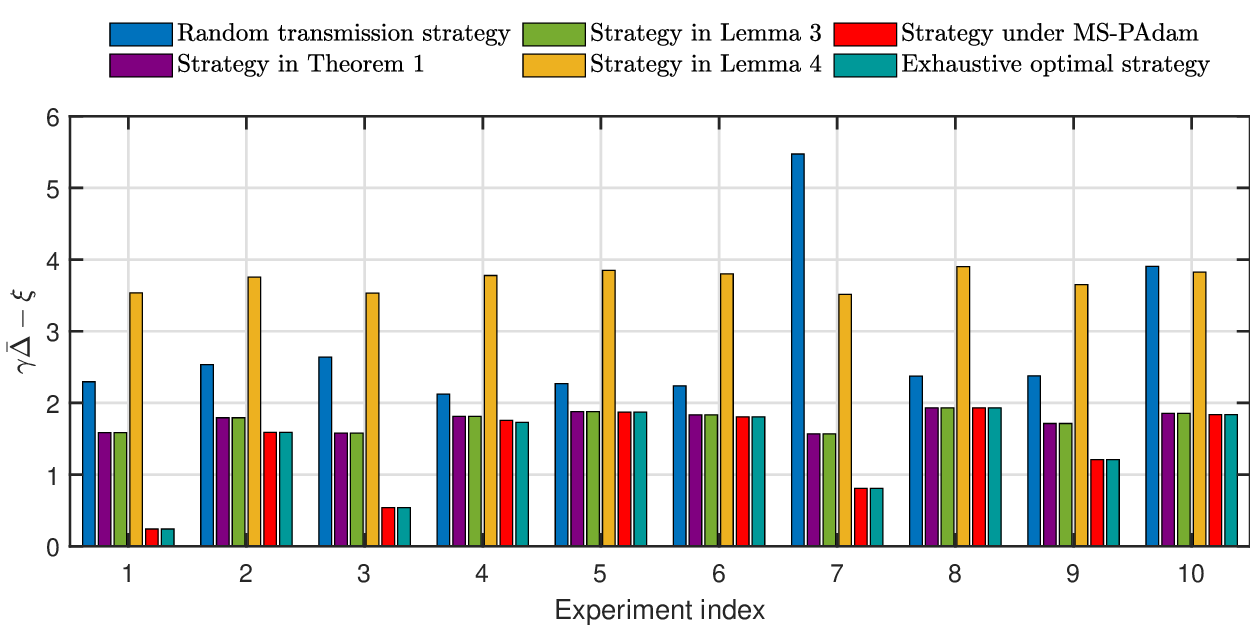}
    \caption{The network performance under different transmission strategies with $n=3$ sensors. The correlation matrix $\textbf{C}$ is randomly generated, with off-diagonal elements $c_{ij}$ chosen from the range [0,0.5].}
    \label{fig:optimal_validate}
\end{figure}

Fig. \ref{fig:optimal_validate} presents three key findings regarding the optimization of $\gamma_{1}\bar{\Delta}-\gamma_{2}\xi$: First, it compares the performance across different transmission strategies, with the exhaustive optimal strategy establishing the theoretical lower bound. Second, it demonstrates that MS-PAdam consistently achieves this performance bound across all ten correlation scenarios. Third, the results confirm that MS-PAdam outperforms all benchmark strategies while converging to the optimal value of $\gamma_{1}\bar{\Delta}-\gamma_{2}\xi$ in every tested scenario.

\begin{figure}[t]
    \centering
    \includegraphics[width=0.48\textwidth]{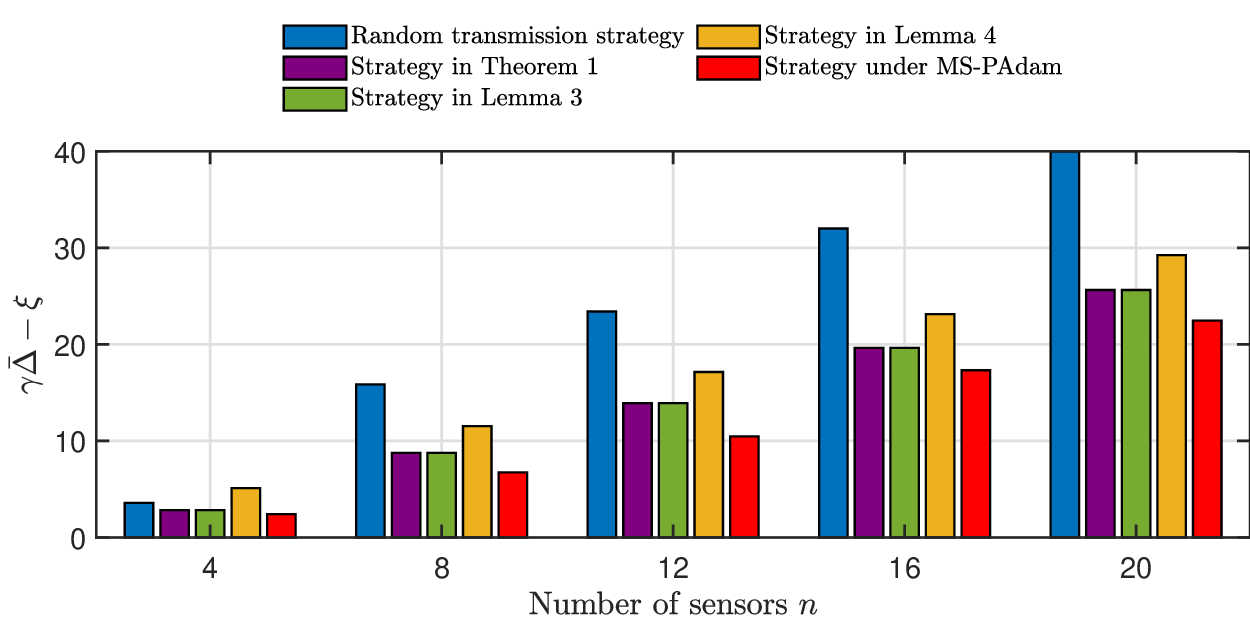}
    \caption{The network performance versus the number of sensors $n$ under different transmission strategies. The correlation matrix $\textbf{C}$ is randomly generated, with non-zero off-diagonal elements $c_{ij}$ selected from the range [0,0.5].}
    \label{fig: global numerical results1}
\end{figure}
\begin{figure}[t]
    \centering
    \includegraphics[width=0.48\textwidth]{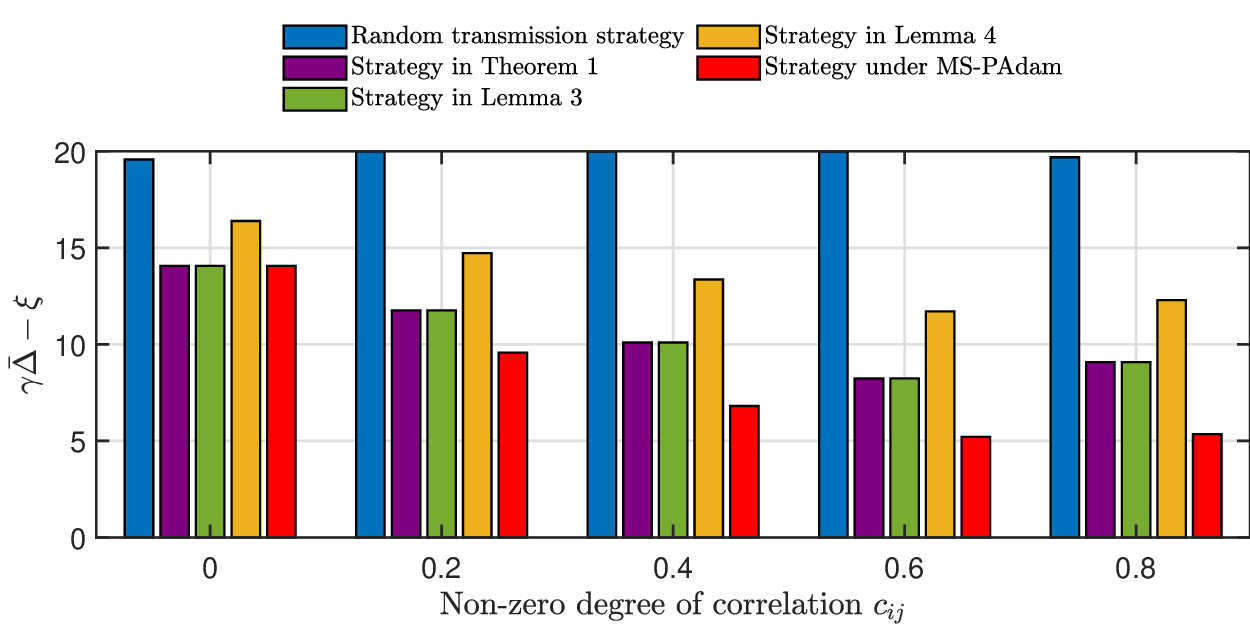}
    \caption{The network performance versus the correlation matrix $\textbf{C}$ under different transmission strategies. The number of sensors $n=10$.}
    \label{fig: global numerical results2}
\end{figure}

\begin{table*}[t]
	\centering
	\caption{The corresponding optimal transmission strategy $\boldsymbol{q}^*$ in Fig. \ref{fig: global numerical results1} with different number of sensors $n$ under MS-PAdam Algorithm.}
	\resizebox{1\linewidth}{!}{
		\begin{tabular}{>{\centering}m{1.5cm}|>{\centering\arraybackslash}m{0.5cm}|>{\centering\arraybackslash}m{0.5cm}|>{\centering\arraybackslash}m{0.5cm}|>{\centering\arraybackslash}m{0.5cm}|>{\centering\arraybackslash}m{0.5cm}|>{\centering\arraybackslash}m{0.5cm}|>{\centering\arraybackslash}m{0.5cm}|>{\centering\arraybackslash}m{0.5cm}|>{\centering\arraybackslash}m{0.5cm}|>{\centering\arraybackslash}m{0.5cm}|>{\centering\arraybackslash}m{0.5cm}|>{\centering\arraybackslash}m{0.5cm}|>{\centering\arraybackslash}m{0.5cm}|>{\centering\arraybackslash}m{0.5cm}|>{\centering\arraybackslash}m{0.5cm}|>{\centering\arraybackslash}m{0.5cm}|>{\centering\arraybackslash}m{0.5cm}|>{\centering\arraybackslash}m{0.5cm}|>{\centering\arraybackslash}m{0.5cm}|>{\centering\arraybackslash}m{0.5cm}
        }\toprule
			optimal solution $\boldsymbol{q}^*$    & $q_1^*$ & $q_2^*$ & $q_3^*$ & $q_4^*$ & $q_5^*$ & $q_6^*$ & $q_7^*$ & $q_8^*$ & $q_9^*$ & $q_{10}^*$ & $q_{11}^*$ & $q_{12}^*$ & $q_{13}^*$ & $q_{14}^*$ & $q_{15}^*$ & $q_{16}^*$ & $q_{17}^*$ & $q_{18}^*$ & $q_{19}^*$ & $q_{20}^*$    \\ \midrule
			1. $n=4$              & 0 & 0 & 0.288 & 0.711 &--&--&--&--&--&--&--&--&--&--&--&--&--&--&--&--      \\ \midrule  
			2. $n=8$              & 0 & 0.136 & 0.143 & 0 & 0.340 & 0 & 0.379 & 0  &--&--&--&--&--&--&--&--&--&--&--&--    \\ \midrule
			3. $n=12$             & 0 & 0 & 0 & 0.511 & 0 & 0 & 0 & 0 & 0 & 0.312 & 0 & 0.177  &--&--&--&--&--&--&--&--         \\ \midrule
			4. $n=16$             & 0 & 0.377 & 0 & 0 & 0 & 0 & 0.019 & 0 & 0 & 0 & 0.255 & 0 & 0.187 & 0 & 0.161 & 0   &--&--&--&--       \\ \midrule 
			5. $n=20$             & 0 & 0 & 0.335 & 0 & 0.297 & 0.111 & 0.106 & 0 & 0 & 0 & 0 & 0.148 & 0 & 0 & 0 & 0 & 0 & 0 & 0 & 0           \\ 
            \bottomrule
	\end{tabular}
}
	\label{table2}
\end{table*}

\begin{table}[t]
	\centering
	\caption{The corresponding optimal transmission probability vector $\boldsymbol{q}^*$ in Fig. \ref{fig: global numerical results2} with different correlated matrix $\mathbf{C}$ under MS-PAdam Algorithm.}
	\resizebox{1\linewidth}{!}{
		\begin{tabular}{>{\centering}m{1.5cm}|>{\centering\arraybackslash}m{0.5cm}|>{\centering\arraybackslash}m{0.5cm}|>{\centering\arraybackslash}m{0.5cm}|>{\centering\arraybackslash}m{0.5cm}|>{\centering\arraybackslash}m{0.5cm}|>{\centering\arraybackslash}m{0.5cm}|>{\centering\arraybackslash}m{0.5cm}|>{\centering\arraybackslash}m{0.5cm}|>{\centering\arraybackslash}m{0.5cm}|>{\centering\arraybackslash}m{0.5cm}}\toprule
			optimal solution $\boldsymbol{q}^*$    & $q_1^*$ & $q_2^*$ & $q_3^*$ & $q_4^*$ & $q_5^*$ & $q_6^*$ & $q_7^*$ & $q_8^*$ & $q_9^*$ & $q_{10}^*$    \\
            \midrule
            1. $c_{ij} = 0$ & 0.099 & 0.099 & 0.098 & 0.099 & 0.098 & 0.099 & 0.099 & 0.099 & 0.099 & 0.099   \\
            2. $c_{ij} = 0.2$  & 0.120 & 0 & 0 & 0 & 0 & 0.168 & 0 & 0.711 & 0 & 0  \\
            3. $c_{ij} = 0.4$  & 0 & 0.487 & 0 & 0.271 & 0 & 0 & 0.24 & 0 & 0 & 0 \\
            4. $c_{ij} = 0.6$  & 0 & 0 & 0 & 0 & 0 & 0 & 0.068 & 0.545 & 0.386 & 0 \\
            5. $c_{ij} = 0.8$  & 0.415 & 0 & 0 & 0 & 0 & 0 & 0.238 & 0 & 0 & 0.345 \\
             \bottomrule
	\end{tabular}
}
	\label{table3}
\end{table}

Fig. \ref{fig: global numerical results1} compares the global network performance optimization results across varying numbers of sensors $n$ under 5 different transmission strategies. In Fig. \ref{fig: global numerical results1}, the correlation matrix $\textbf{C}$ is randomly generated, with off-diagonal elements $c_{ij}$ (i.e., degree of correlation) selected from the range [0,0.5]. From Fig. \ref{fig: global numerical results1} we can conclude that when the network scale is small (n=4), all strategies show comparable performance due to limited network competition. However, as the network scale increases, MS-PAdam demonstrates significantly better performance than other benchmark strategies. This performance gap highlights MS-PAdam's ability to leverage inter-sensor correlations to optimize both average AoI and energy efficiency in large-scale networks..

Fig. \ref{fig: global numerical results2} presents a comparison of the global network performance optimization versus the \textit{non-zero degree of correlation, defined as the values of the non-zero off-diagonal elements $c_{ij}$ in $\mathbf{C}$} under 5 different transmission strategies.
In Fig. \ref{fig: global numerical results2}, when all off-diagonal elements $c_{ij}=0$, the correlation structure $\mathbf{C}$ in this case is equivalent to the Identity matrix $\mathbf{I}$, indicating that all sensors are independent. In this configuration, the MS-PAdam algorithm achieves network performance that asymptotically approaches the homogeneous optimal performance established in Theorem \ref{theorem 1}.
Consequently, both Theorem \ref{theorem 1} and the first row of Table \ref{table3} demonstrate that when there is no correlation between sensors, the optimal transmission probability vector configuration of the network is equivalent to the homogeneous network, i.e., $\boldsymbol{q}^* \in [q_E^*, q_A^*]^n$. However, under such a configuration, competition among sensors in the network becomes most intense, as the distributed access behavior and each node shares the same transmission probability.

In contrast to the independent case, as discussed below, the situation changes when considering the correlation between the observations of sensors. Fig. \ref{fig: global numerical results2} also presents the optimized network performance under correlation setting, i.e., when the non-zero off-diagonal elements $c_{ij}$ in correlation matrix $\mathbf{C}$ are not all zero. 
Fig. \ref{fig: global numerical results2} demonstrates that in the correlated random access network, the network average and energy efficiency in both homogeneous and heterogeneous cases can be further optimized. As the non-zero degree of correlation increases, the MS-PAdam algorithm optimizes the network performance, $\gamma_{1} \bar{\Delta}^* - \gamma_{2}\xi^*$, by 32\%, 52\%, 63\%, and 62\% compared to the optimal cases in the independent network. 
Additionally, the gap between the homogeneous optimal strategy and the MS-PAdam strategy increases, indicating the favorable performance of the MS-PAdam method.

Furthermore, both Table \ref{table2} and Table \ref{table3} demonstrate that collisions are significantly reduced in the correlated random access network. These are interesting network behaviors. In Table \ref{table2}, only no more than 5 active sensors remain accessing the channel is optimal, while in Table \ref{table3}, more than 6 sensors prefer to yield the channel to other nodes for transmitting status updates rather than transmitting themselves. This preference increases with the degree of correlation, as the remaining nodes can effectively assist in relaying their neighbors' status updates to the receiver. As a result, collisions caused by distributed access among sensors can be reduced by at least 50\%. Here, we define the contention reduction ratio as the ratio of the number of sensors that choose not to compete the channel to the total number of sensors.

Additionally, Fig. \ref{fig: global numerical results2} demonstrates that impractical transmission strategies (e.g., random transmission strategy, homogeneous age-optimal strategy in Lemma \ref{lemma 3}, and homogeneous energy-efficiency optimal strategy in Lemma \ref{lemma 4}) significantly degrade network performance, even in the presence of strong correlations between sensors. In contrast, our proposed MS-PAdam method achieves a more seasonable transmission probability configuration while ensuring optimal network average AoI performance.
Our analysis systematically investigates the role of correlation structures $\mathbf{C}$ and and sensor density $n$ in network optimization, showing how spatial correlation can simultaneously improve AoI performance and energy efficiency. The study further quantifies these effects on transmission strategies, demonstrating that correlation-aware optimization significantly decreases collision probabilities in random access scenarios. These findings demonstrate how inherent correlation patterns naturally coordinate sensor transmissions, yielding three key benefits: improved information freshness, enhanced energy efficiency, and reduced contention-induced packet loss.

\section{Conclusion} \label{Conclusion}
This paper characterizes and optimizes the AoI and energy efficiency in heterogeneous random access networks with correlated sensors, where each sensor is associated with a distinct transmission probability and exhibits correlation with other sensors. Specifically, we derive a closed-form expression for energy efficiency, model the AoI dynamics using a discrete-time Markov process, and obtain the long-term average AoI for each sensor. Furthermore, assuming equal transmission probabilities across all sensors, we derive the optimal transmission strategies for the age-optimal, energy-efficiency-optimal, and joint age-energy-optimal cases. respectively.
Additionally, when each sensor is assigned a distinct transmission probability, a gradient-based algorithm, MS-PAdam, is proposed to jointly optimize the network average AoI and energy efficiency, thereby yielding the optimal transmission strategy. The proposed algorithm exhibits rapid and reliable convergence to the optimal network performance while simultaneously delivering the corresponding transmission strategy.

Analyses show that, under the assumption of equal transmission probabilities across all sensors, the optimal transmission probability lies within the interval $q^* \in [q_E^*, q_A^*]$, where $q_E^*$ denotes the energy-efficiency-optimal strategy and $q_A^*$ the age-optimal strategy. Moreover, both theoretical results and simulations indicate that when all sensors are independent, i.e., when $\mathbf{C}$ is an identity matrix, the optimal transmission strategy coincides with that of the homogeneous network, as established in Theorem \ref{theorem 1}. In contrast, in the presence of correlations among sensors, the network can sustain satisfactory AoI performance and energy efficiency even with reduced transmission frequency or complete abstention from transmission. Numerical results further reveal that the proposed MS-PAdam method achieves a balanced transmission strategy, mitigating performance degradation caused by severe contention in high-density distributed access networks while preserving near-optimal network AoI performance. Finally, our analysis and the proposed algorithm demonstrate that in heterogeneous random access networks with correlated status updates, sensor competition is significantly reduced and overall network performance is substantially enhanced.

\appendices
\section{Proof of Lemma \ref{Lemma 1}} \label{proof of lemma 1}
According to Fig. \ref{AoI_Markovdynamics}, we have:
\begin{equation}\label{appendixLemma_eq1}
\begin{cases}
    \pi_{i}^{1} =  p_{i}^{r}\cdot \sum_{k=1}^{\hat{\Delta}} \pi_{i}^{k},\\
    \pi_{i}^{k} = (1-p_{i}^{r}) \cdot \pi_{i}^{k-1} \text{  for }k=2,3,\ldots,\hat{\Delta}-1,\\
    \pi_{i}^{\hat{\Delta}} = (1-p_{i}^{r}) \cdot ( \pi_{i}^{\hat{\Delta}-1} + \pi_{i}^{\hat{\Delta}})\\
\end{cases}
\end{equation}
Note that the following equation always holds:
\begin{equation}\label{appendixLemma_eq2}
    \sum_{k=1}^{\hat{\Delta}} \pi_{i}^{k} = 1.
\end{equation}
where $p_{i}^{r}$ is given by \eqref{AoI_reset_prob}.
Combining with \eqref{appendixLemma_eq1} and \eqref{appendixLemma_eq2}, the steady-state distribution of the Markov chain can be obtained.
We have the following equation following the definition of the Long-term average AoI: 
\begin{equation}
\begin{aligned}
       \bar{\Delta}_i 
         &  \overset{\triangle}{=} \lim_{T\to\infty}\sup\frac{1}{T}\sum_{t=0}^{T-1}\Delta_i(t) =\sum_{k=1}^{\hat{\Delta}}\pi_i^k \cdot k \\
         & =  \sum_{k=1}^{\hat{\Delta}-1}(1-p_{i}^{r})^{k-1}p_{i}^{r}k + (1-p_{i}^{r})^{\hat{\Delta}-1}\hat{\Delta}\\
         & = \frac{1-(1-p_{i}^{r})^{\hat{\Delta}}}{p_{i}^{r}}\\
\end{aligned}
\end{equation}
where $p_{i}^{r}$ is given by \eqref{AoI_reset_prob}. Therefore, we derive the expression of the long-term average AoI of sensor $i$.
The proof of Lemma \ref{Lemma 1} is completed.

\section{Proof of Lemma \ref{Lemma 2}} \label{proof of lemma 2}
Given the finite amount of energy $E$, the expected lifetime can be expressed as
\begin{equation}
    \tau_i = \frac{E}{q_i P_{T,i} + (1-q_i) P_I},
\end{equation}
The network throughput of each sensor is defined as the probability of successful transmission of status updates during one time slot. Combining with the spatial correlation among sensors, the throughput of the sensor $i$ can be expressed as
\begin{equation}
    \lambda_{out,i} = \sum_{j\in \mathcal{N}} p_j q_j c_{ji}.
\end{equation}
Since the lifetime throughput is a product of the throughput and the lifetime of the sensor \cite{Anshan2024EE}, the lifetime throughput of sensor $i$ is given by
\begin{equation}
    U_i = \frac{E\sum_{j\in \mathcal{N}} p_j q_j c_{ji}}{q_i P_{T,i} + (1-q_i) P_I}.
\end{equation}

\section{Proof of Lemma \ref{lemma 3}}\label{proof of lemma 3}

According to \eqref{average AoI for one node}, we know that the long-term average AoI $\bar{\Delta}_i$ of sensor $i$ is given by:
\begin{equation}
    \bar{\Delta}_i =\frac{1-(1-p_{i}^{r})^{\hat{\Delta}}}{p_{i}^{r}} \underset{p_{i}^{r} \in (0,1)}{\overset{\hat{\Delta}\gg 1}{\approx}} \frac{1-\exp \left\{-{\hat{\Delta}p_{i}^{r}} \right\}}{p_{i}^{r}}
\end{equation}
For simplicity of expression, below we use the approximated expression, i.e.,
\begin{equation}
    \bar{\Delta}_i = \frac{1-\exp \left\{-{\hat{\Delta}p_{i}^{r}} \right\}}{p_{i}^{r}}.
\end{equation}
Note that $\bar{\Delta}_i$ is the function of the AoI reset probability of sensor $i$, i.e., $p_{i}^{r}$. Thus we can take the first and second derivatives of $p_{i}^{r}$, which can be written as follows:
\begin{equation}
   \bar{\Delta}_i' = \frac{\partial \bar{\Delta}_i}{\partial p_{i}^{r}} = \frac{p_{i}^{r}\hat{\Delta}\exp \left\{-{\hat{\Delta}p_{i}^{r}}\right\}+\exp \left\{-{\hat{\Delta}p_{i}^{r}}\right\}-1}{{p_{i}^{r}}^2}.
\end{equation}
Since ${p_{i}^{r}}^2 > 0$, we only need to consider the numerator. Let 
\begin{equation}
    f(p_{i}^{r}) =p_{i}^{r}\hat{\Delta}\exp \left\{-{\hat{\Delta}p_{i}^{r}}\right\}+\exp \left\{-{\hat{\Delta}p_{i}^{r}}\right\}-1,
\end{equation}
then 
\begin{equation}
   f'(p_{i}^{r}) = \frac{\partial f(p_{i}^{r})}{\partial p_{i}^{r}} = -\hat{\Delta}p_{i}^{r}\exp \left\{-{\hat{\Delta}p_{i}^{r}}\right\}.
\end{equation}
Note that $0\leq p_{i}^{r} = p q\sum_{j=1}^n c_{ji} \leq np q\leq \exp(-1)$ and $\hat{\Delta}\geq1$, we can obtain that $f'(p_{i}^{r})\leq 0$ always holds. Thus we can obtain that $\bar{\Delta}_i' = \frac{\partial \bar{\Delta}_i}{\partial p_{i}^{r}}$ decreases monotonically as $p_{i}^{r}$ increases.

Therefore, the maximum value of $\bar{\Delta}_i' = \frac{\partial \bar{\Delta}_i}{\partial p_{i}^{r}}$ should be obtained when $p_{i}^{r}$ reaches its minimum value. Since we have $0\leq p_{i}^{r} = p q\sum_{j=1}^n c_{ji} \leq np q\leq \exp(-1)$, the maximum value of $\bar{\Delta}_i'$ is given by:
\begin{equation}
    \lim_{p_{i}^{r}=0} \bar{\Delta}_i'~~{\overset{\text{L'Hôpital's Rule}}{=}} ~~ \frac{\hat{\Delta}-\hat{\Delta}^2}{2}\leq 0.
\end{equation}

As a result, we have proved that $\bar{\Delta}_i' \leq 0$ always hold for $\forall{p_{i}^{r}}$. Then we can further obtain that $\bar{\Delta}_i $ decreases monotonically as $p_{i}^{r}$ increases, indicating that for any sensor $i$, its correlated AoI is minimized when its corresponding $p_{i}^{r}$ is maximized.

Furthermore, in homogeneous scenario, $p_{i}^{r} = p q\sum_{j=1}^n c_{ji} $, and $p = \exp(-nq)$ with a large $n$, we can rewrite the AoI reset probability of sensor $i$ as follows:
\begin{equation}
    p_{i}^{r} = q\exp(-nq)\sum_{j=1}^n c_{ji},
\end{equation}
where $\sum_{j=1}^n c_{ji}$ is a quantity independent of the transmission probability $q$ of the sensor. When the correlation matrix $\mathbf{C}$ is given, $\sum_{j=1}^n c_{ji}$ is a constant. Therefore, it is easy to get the optimal transmission probability $q_A^* = \frac{1}{n}$ to maximize the AoI reset probability of sensor $i$ by taking the derivative of $p_{i}^{r}$ to the transmission probability $q$. 
As a result, the optimal AoI reset probability of sensor $i$ can be obtained as follows:
\begin{equation}
    p_{i}^{r*} = \frac{\exp(-1)\sum_{j=1}^n c_{ji}}{n}.
\end{equation}
Then we complete the proof.

\section{Proof of Lemma \ref{lemma 4}}\label{proof of lemma 4}
Given constraints $q_i = q_j $ for $\forall i,j \in \mathcal{N}$ and a large $n$, the energy efficiency of each sensor can be written as
\begin{equation}
    \xi_i = \frac{pq\sum_{j\in \mathcal{N}}  c_{ji}}{P_I+q(P_T-P_I)} =\frac{q\exp(-nq)\sum_{j\in \mathcal{N}}  c_{ji}}{P_I+q(P_T-P_I)} .
\end{equation}
The network energy efficiency is given by $\xi = \sum_{i\in\mathcal{N}} \xi_i.$ $q_E^*$ can be obtained by solving $\frac{\partial\xi}{\partial q} = 0$. Note that $\xi$ increases monotonically when $q<q_E^*$ and decreases monotonically when $q>q_E^*$, therefore, $\xi$ is maximize when $q =q_E^*$.

\section{Proof of Theorem \ref{theorem 1}} \label{proof of theorem 1}
Let $J(q) = \gamma_{1}\bar{\Delta}(q) - \gamma_{2}\xi(q)$. 
Combining with Lemma \ref{lemma 3} and Lemma \ref{lemma 4}, we have $\bar{\Delta}$ decreases monotonically when $q<q_A^*$ and increases monotonically when $q>q_A^*$, while $\xi$ increases monotonically when $q<q_E^*$ and decreases monotonically when $q>q_E^*$. Additionally, it can be easily proven that $q_E^* \leq q_A^*$ always holds if $P_T \geq P_I$. Consequently, it can be proven that $J(q) \gets \gamma_{1}\bar{\Delta}(q) - \gamma_{2}\xi(q)$ decreases monotonically when $q<q_E^*$ and increases monotonically when $q>q_A^*$. Thus, the optimal $q^*$ that minimizes $J(q)$ lies within the interval $[q_E^*,q_A^*]$.

\bibliographystyle{IEEEtran}
\bibliography{IEEEabrv,Fn_Ref}

\end{document}